\documentclass[aps,pre,twocolumn,showpacs]{revtex4-1}

\usepackage{amsfonts,amssymb,amsmath}

\usepackage{times}
\usepackage{graphicx}
\usepackage{bm}
\usepackage{color}

    \newcommand{\bee}{\begin{equation}}
\newcommand{\eee}{\end{equation}}
\newcommand{\ba}{\begin{eqnarray}}
\newcommand{\ea}{\end{eqnarray}}

\newcommand{\ket}[1]{\vert #1 \rangle}
\newcommand{\bra}[1]{\langle #1 \vert}

\begin{document}
\title{\bf  Effect of particle collisions in dense suspension flows}
\author{Gustavo D\"uring${}^{1}$, Edan Lerner${}^{2}$ and Matthieu Wyart${}^{3}$}

\affiliation {
${}^1$ Facultad de F\'isica, Pontificia Universidad Cat\'olica de Chile, Casilla 306, Santiago, Chile\\
${}^2$ Institute for Theoretical Physics, Institute of Physics, University of Amsterdam, Science Park 904, 1098 XH Amsterdam, The Netherlands\\
${}^3$Institute of Theoretical Physics, \'Ecole Polytechnique F\'ed\'erale de Lausanne, CH-1015 Lausanne, Switzerland\\
}
\date{\today}
\begin{abstract}
We study non-local effects associated with  particle collisions in dense suspension flows, in the context of the affine solvent model
known to capture various aspects of the jamming transition. We show that an individual collision changes significantly the velocity field on
a characteristic volume $\Omega_c\sim 1/\delta z$ that diverges as jamming is approached, where $\delta z$ is the deficit in coordination number
 required to jam the system. Such an event also affects the contact forces between particles on that same volume $\Omega_c$, but this change is modest in relative terms, of order  $f_{coll}\sim \bar{f}^{0.8}$, where $\bar{f}$ is the typical contact force scale. We then show that the requirement that coordination is stationary (such that a collision has a finite probability to open one contact elsewhere in the system)   yields the scaling of the viscosity (or equivalently the viscous number) with coordination deficit $\delta z$.   The same scaling result was derived in [E.~DeGiuli, G.~D\"uring, E.~Lerner, and M.~Wyart, Phys.~Rev.~E {\bf 91}, 062206 (2015)] via different arguments making an additional assumption.  The present approach gives a mechanistic justification as to why the correct finite size scaling volume behaves as $1/\delta z$, and can be used to recover a marginality condition known to characterize the distributions of contact forces and gaps in jammed packings.

\end{abstract}

\maketitle

\section{Introduction} Suspensions are complex fluids consisting of solid particles immersed in a viscous liquid.  The presence of solid particles affects flows, especially when the concentration of particles or the so called packing fraction $\phi$ becomes large. In the dilute limit  Einstein proved that the presence of particles leads to a linear increase of the viscosity $\eta$ with $\phi$ \cite{Einstein06}. However, the dilute regime breaks down upon densification as steric-hindrance effects become dominant. At larger packing fractions \cite{Brown09,Boyer11}  the viscosity even diverges at the jamming point $\phi_c$ where the suspension jams into an amorphous solid. Critical exponents governing the rheology of dense suspensions as well as a diverging correlation length scale have been observed in experiments \cite{Brown09,Boyer11,Pouliquen04,Lespiat11,Nordstrom10} and in numerical models \cite{Durian95,Olsson07,Olsson11,Hatano08,Lerner12a,Andreotti12,Trulsson12,Vagberg14b,Ikeda12,Wang15}. 

In the context of frictionless particles, we have proposed together with others a microscopic description that predicts both the explosion of the correlation length of velocity fluctuations \cite{During14} as well as the critical rheological properties of both over-damped and inertial flows \cite{DeGiuli15a}. This approach has receive recent numerical \cite{Vagberg15,DeGiuli15a} and empirical \cite{Dagois15} support. However, it makes an  assumption on the nature of  flowing configurations, thought to be similar to slightly perturbed  jammed configurations. It also predicts that the finite size volume scales as $1/\delta z$, which differs (except in two dimensions) from the naive estimate $\xi^d$, where $\xi\sim 1/\sqrt{\delta z}$ is the main length scale on which velocity correlations decay. % Finally, this approach is limited to stationary flow, and does not deal for example with a shear reversal \cite{Blanc11}. 

In this work, we show how to recover some of the scaling results of \cite{DeGiuli15a}  with less assumptions in the framework of the Affine Solvent Model (ASM), where the viscous damping neglects hydrodynamic interactions \cite{Durian95,Olsson07} and particles are perfectly hard \cite{Lerner12a,Lerner13}. Our work is based on a detailed description of the effect of an individual collision between particles on the velocity field and contact forces, which will presumably be of value to understand how perturbations (such as a shear reversal \cite{Blanc11}) affect structure and flow. The observation that the mean contact number must not evolve in average, implies that a collision (which forms a new contact) must have a finite probability to open exactly one contact, which yields a scaling relation between coordination and viscosity. Our work justifies further why the characteristic finite-size volume  varies as $1/\delta z$ \cite{DeGiuli15a}, as this is precisely the characteristic volume over which the mechanical effect of a collision extends.

\begin{figure*}
\includegraphics[width=1\textwidth,trim=7.8cm 3.5cm 4.5cm 3.cm, clip=true ]{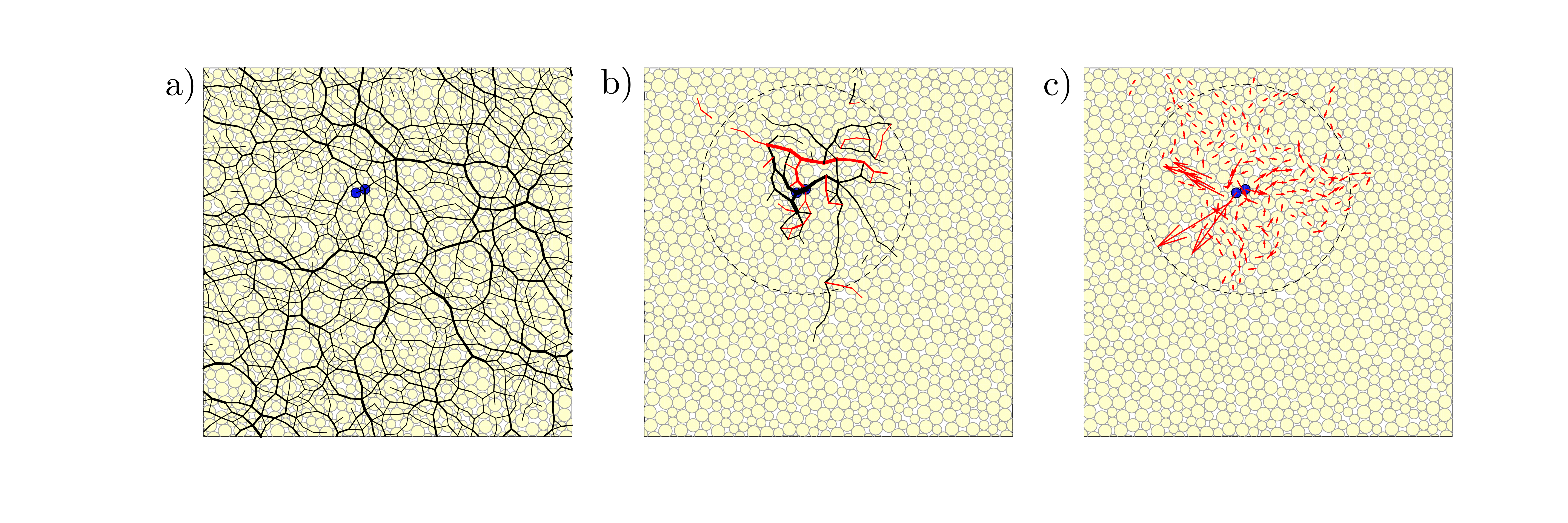} 
\caption{Snapshot a of suspension flowing under simple shear using the ASM at the instant of a collision between two particles (in blue). {\bf a)} Black lines represent the contact network; the width of the lines are proportional to the magnitude of contact forces $f$ immediately after the collision.  {\bf b)} The width of the lines connecting centers of particles are proportional to the magnitude of the \emph{instantaneous variations} in contact forces induced by the collision. Red (black) lines correspond to negative (positive) variations in the contact forces. The dashed circle is a visualization of the typical volume $\Omega_c$ as defined in the text.  {\bf c)} The vector field represents the \emph{instantaneous variations} in the particles' velocity induced by the collision. The dashed circle is a visualization of the typical volume $\Omega_v$ as defined in the text.}
\label{collfig1}
\end{figure*}

 \section{The affine solvent model}  
  
The ASM is an idealized suspension model which has been shown to have at least qualitatively good agreement with the rheology of dense suspension flows \cite{Lerner12a}. The model considers $N$ frictionless hard spherical particles in a volume $\Omega$ immersed in a viscous fluid of viscosity $\eta_0$, and hydrodynamic interactions are neglected. The viscous fluid act as a carrier with a velocity profile  $\vec{V}^f(\vec{R})$ which depends on the spatial position $\vec{R}$. For the sake of simplicity, we shall consider simple shear flow in the $x,y$-plane at constant volume  with a strain rate $\dot{\gamma}$, hence $\vec{V}^f(\vec{R})=\dot{\gamma}y \hat{x}$. The ASM can be easily extended to flows under constant confining pressure instead of constant volume \cite{Lerner13}. However, the bulk properties derived in this paper remain unchanged between the two ensembles in the thermodynamic limit.

We consider overdamped dynamics such that the viscous fluid induces a Stokes' drag force proportional to the velocity difference between the particles velocity $\vec{V}_k$ and the fluid velocity $\vec{V}^f(\vec{R}_k)$, where $\vec{R}_{k}$ is the position of the $k^{\mbox{\scriptsize th}}$ particle. Hence, the drag force is written as 
\begin{equation}
\vec{F}_k=-\eta_0 r_0(\vec{V}_k-\vec{V}^f(\vec{R}_k)),
\label{dragForce}
\end{equation}
 where $r_0$ is the mean particle diameter. The absence of inertia implies the formation of persistent contacts which form a network as shown in Fig. \ref{collfig1}a. The repulsive contact force between two hard spheres will be taken to be positive. The total number of contacts $N_c$ defines the coordination number $z=2 N_c/N$. In what follows, contacts will be labeled with greek letters, e.g.~the pair of particles $i$ and $k$ in contact will be labeled as $\beta$, with the contact force $f_\beta$.

The relative  radial velocity between particles $i$ and $k$ is given by
\bee
v_{ik}= (\vec{V}_k-\vec{V}_i)\cdot\vec{n}_{ik},
\label{radialVelocity}
\eee
where the unit vector $\vec{n}_{ik}$ points along the difference $\vec{R}_k-\vec{R}_i$. A positive value of $v_{ik}$ represents pairs of particles moving apart from each other. Hard particles cannot overlap, thus if $i$ is in contact with $k$, the relative radial velocity $v_{ik}$ must be zero. The resulting set of $N_c$ equations (\ref{radialVelocity}) for particles in contact are linear in the $N$ particle velocities and can be written in a matrix form as
\bee
\mathcal{S}|V\rangle=0
\label{hc1}
\eee
where the  operator $\mathcal{S}$ depends only on the unit vectors $\vec{n}_{ik}$ \cite{Lerner13a}. The vector $|V\rangle$ of dimension $ND$ represents the velocity field of the entire system, i.e $\langle i|V\rangle=\vec{V}_i$. Notice that $\mathcal{S}$ is a non-square matrix of dimension $N_c\times ND$. The velocity profile of the fluid can also be written in compact notation as $|V^f\rangle$, where $\langle k|V^f\rangle=\vec{V}^f(\vec{R}_k)$.

From the expression of the drag force, the requirement that forces are balanced, and the non-overlap  constraints (\ref{hc1}) one can compute the instantaneous contacts forces \cite{Lerner12}:
\bee
 |f\rangle=-\eta_0 r_0\dot{\gamma}\mathcal{N}^{-1}|\gamma\rangle\,
\label{contactForce}
\eee 
where  $\mathcal{N}=\mathcal{S}\mathcal{S}^t$  and  $|\gamma\rangle=\mathcal{S}|V^f\rangle/\dot{\gamma}$. $|\gamma\rangle$ is a non-singular vector of dimension $N_c$ which indicates the  imposed deformation mode  (see the supplementary information for details and for a derivation of Eq.\ref{contactForce}). In what follows, lowercase vectors correspond to contact-space vectors of dimension $N_c$, e.g. $f_\beta=\langle \beta|f\rangle$, while uppercase vectors belong to particle space, of dimension $ND$. The $\mathcal{N}$ matrix depends solely on the geometry of the network formed by the contacts, and allows us to determine the rheological properties  of the suspension.

The evolution of the system is determined by the velocity field:
\bee
|V\rangle=\frac{\mathcal{S}^t|f\rangle}{\eta_0 r_0 }+|V^f\rangle.
\label{Velocity}
\eee
The shear stress and the pressure are defined as
\begin{eqnarray}
\sigma\equiv-\frac{\langle \gamma|f\rangle}{\Omega}\quad  \& \quad p\equiv \frac{\langle r|f\rangle}{D\Omega}
\label{sigma}
\end{eqnarray}
respectively, where $r_\beta\equiv\langle \beta|r\rangle$ is the distance between the particles that form the contact $\beta$. Obviously one has  $$\bar{f}\sim r^{D-1}_0 p  = \eta_0 \dot{\gamma} r^{D-1}_0 \, \mathcal{J}^{-1} $$ where $ \mathcal{J}=\frac{\eta_0\dot{\gamma}}{p}$ is the  dimensionless viscous number \cite{Boyer11}.

Fluctuations of the velocity with respect to the affine flow are given by the non-affine velocity $|V_\text{n.a}\rangle= |V\rangle- |V^f\rangle$, with a mean square value of which must follow according to  Eqs.(\ref{contactForce},\ref{Velocity},\ref{sigma}) $V^2_\text{n.a}\sim~\frac{\dot{\gamma}^2 r_0^D}{\Omega}\langle \gamma|\mathcal{N}^{-1}|\gamma\rangle=\sigma \dot{\gamma} r_0^{D-1}/\eta_0$. The latter equation simply corresponds to the balance of power injected and power dissipated by the viscous damping \cite{Lerner12a,Andreotti12}.

In addition, the friction $\mu=\frac{\sigma}{p}$ is known to remain finite at the jamming point, thus as jamming is approached one has:
\bee
p\sim \sigma\sim \eta_0 V^2_\text{n.a}/\dot{\gamma} r_0^{D-1}.
\label{Vna}
\eee
For sake of simplicity  the typical diameter $r_0$, the viscosity $\eta_0$ and the strain rate $\dot{\gamma}$ will be set to unity in what follows. Therefore, the viscous number controlling the rheology reads $\mathcal{J}=1/p$ and either $\mathcal{J}^{-1}$ or $p$ can  be used interchangeably.

\section{Stationarity condition}

We now discuss the constraint resulting from the fact that in the steady flow state, the average number of contacts (and hence also the coordination $z$) reaches a stationary value. We observe that most contact openings occur due to collisions. Thus in average, when a collision occurs and a new contact is formed, another contact must open.  

\begin{figure}[h]
\includegraphics[width=1.1\columnwidth,trim=2cm 0cm 0cm 0cm, clip=true ]{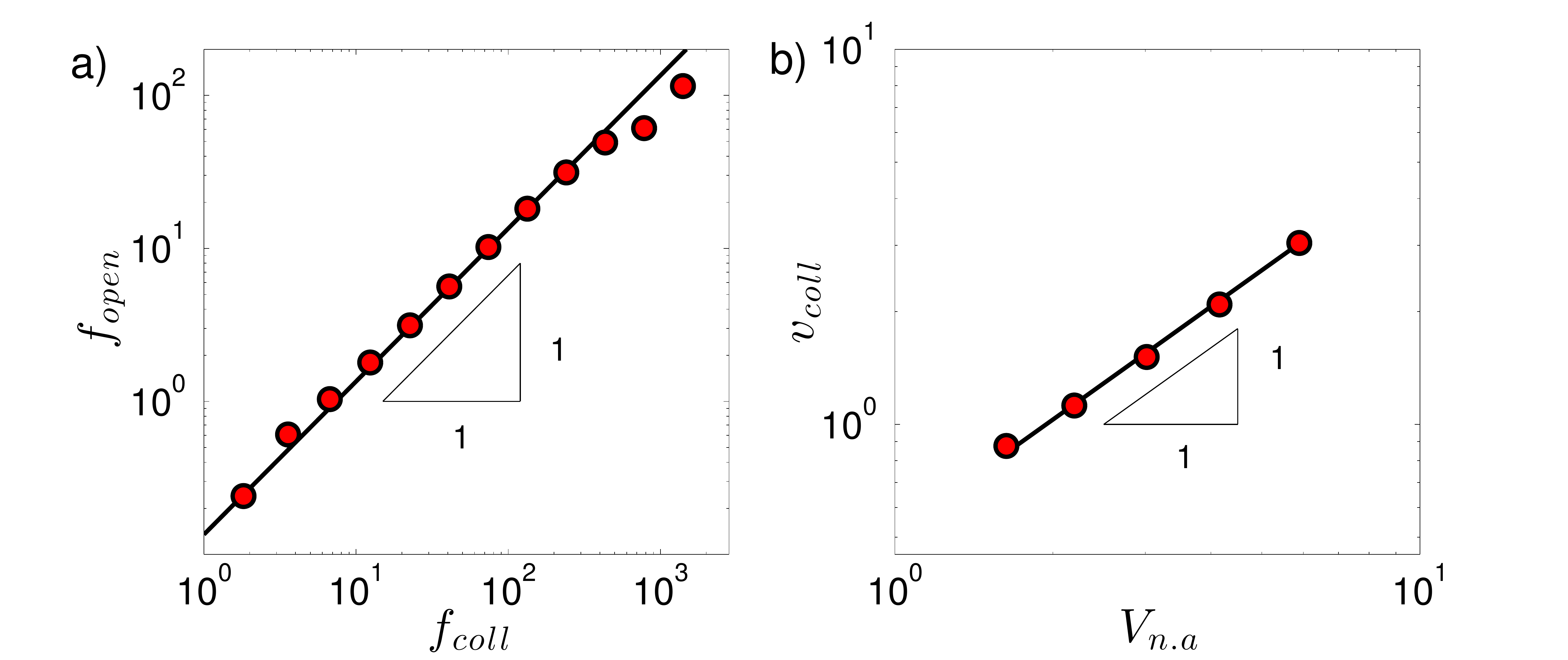} 
\caption{Data from simulations of the ASM in three dimensions, for systems of $N=2000$ particles, and pressures ranging between $10$ and $1000$. Rattlers have been removed from the analysis, see \cite{Lerner13} for details about the procedure of rattlers removal. {\bf a)} Forces $f_\text{open}$ in contacts that open due to a collision, just before the collision takes place, {\it vs} the force in the newly created contact $f_\text{coll}$. {\bf b)} relative radial velocity $v_\text{coll}$ of pairs of colliding particles just before a collision takes place, {\it vs} the mean non-affine velocity of the particles $V_\text{n.a}$, taken over all particles in the system. 
}
\label{collfig}
\end{figure}

To estimate the probability that a contact opens, we must estimate  the forces involved in contact formation and destruction.
A collision between two particles generates a new contact with a force that we denote $f_\text{coll}$. This collision induces a discontinuous change in the surrounding contact forces; the difference  $\Delta f_\beta=f^{\text{after}}_\beta-f^{\text{before}}_\beta$ is displayed in Fig.\ref{collfig1}b. Essentially, forming a contact is analogous to imposing a localize dipolar force on a floppy material, a problem we have studied in detail in \cite{During14,During13}. In an isostatic system ($\delta z=0$), a dipole of amplitude $f_\text{coll}$ would change all forces in the system by $\Delta f_\beta\sim f_\text{coll}$. 
However, in a floppy system ($\delta z > 0$), the amplitude $\Delta f_\beta$ is of order of $f_\text{coll}$ only in the vicinity of the dipole, and eventually decays exponentially (see more on that below).  Therefore, we can define the correlation volume $\Omega_c\equiv \left(\sum_\beta \Delta f^2_\beta\right)/f^2_{\text{coll}}$ as the volume inside which the magnitude of $\Delta f_\beta$ is of the order of $f_{\text{coll}}$. The change in each force $\Delta f_\beta$ can be positive or negative.

At the instance of a collision, the contacts that have a finite probability to open due to the collision are those that reside inside a volume $\Omega_c$ around the collision location, and whose force $f_\text{open}$ before the collision was of order $f_{\text{coll}}$.
 Thus, $f_\text{open}$ is expected to scale as $f_\text{coll}$, as we confirm in Fig.\ref{collfig}a. 
It is clear from this argument that if $\Omega_c \gg1$ (which turns out to be true near jamming, see below), then the collisional force $f_{\text{coll}}$ must be much smaller than the pressure, the latter sets the scale of typical contact forces. Otherwise, many contacts inside the volume $\Omega_c$ would open upon a typical collision event, which would violate the stationarity of the mean coordination. We thus conclude that  the force of the opened contact $f_\text{open}$ must scale as the \emph{weakest force} in the volume $\Omega_c$ \footnote[1]{this argument is somewhat more subtle, because some contacts that carry a weak force are mechanically isolated and are thus insensitive to collisions. Only the contacts mechanically coupled to the rest of the system have to be considered in this argument,  see \cite{Lerner13a}.}.  This leads to the scaling relation:
 \bee
f_{\text{min}}\equiv \min_{\in\Omega_c}{f}\sim f_{\text{coll}}.
 \label{sca1}
 \eee
In what follows we compute  $f_{\text{coll}}$, the volume $\Omega_c$ and $f_{\text{min}}$ to extract a useful scaling relation from Eq.(\ref{sca1}).
\section{Collisional force in the ASM framework}

Pairs of particles that are on course to collide do not behave differently than any other pair of particles, thus the relative velocity with which they collide, referred to in what follows as the collisional velocity $v_\text{coll}$, must scale as the velocity fluctuations $V_\text{n.a}$. This fact is confirmed numerically in Fig.\ref{collfig}b. From (\ref{Vna}) one gets:
\begin{equation}
v_{coll}\sim \sqrt{p}.
\label{vcoll}
\end{equation}

When a collision takes place, the radial relative velocity between the colliding particles jumps discontinuously from $v_\text{coll}$ to $0$ (since the paricles' velocities must respect the constraint that hard particles cannot overlap), while the force in the contact formed jumps from $0$ to some $f_\text{coll}$. This discontinuity of the force in the newly formed contact causes a sudden change in the entire force field $\Delta f_\beta$. A collision can open new contacts with a finite probability. However, to estimate the effect of a collision on the force network, we may assume that no contacts open, as this simplification turns out not to modify our estimates.

%{\color{red}(maybe start with defining $V_a$ etc.?)}
The operation of $\mathcal{S}_a$ (defined on the post-collision contact network) on the post-collision velocities $\ket{V_a}$ is zero by construction, since the relative radial velocities for particles in contact vanish.  However, if $\mathcal{S}_a$ operates on the pre-collision velocities $\ket{V_b}$, one obtains $\mathcal{S}_a|V_b\rangle= v_\text{coll}|\alpha\rangle$, where $\alpha$ is the contact created at the collision. Replacing the constrain (\ref{hc1}) by this relation, one obtains the pre-collision instantaneous response  in terms of the contact network after the collision (see supplementary information for details). The pre-collision contact forces are then given by $\ket{f_b}= -\mathcal{N}_a^{-1}\ket{\gamma}+\mathcal{N}_a^{-1}\ket{\alpha}v_\text{coll},$ where $\mathcal{N}_a\equiv \mathcal{S}_a\mathcal{S}^t_a$ and the first term on the right hand side of the equation corresponds to the post-collision forces $\ket{f_a}$ defined in Eq.(\ref{contactForce}). Thus, the change in the contact forces is
\bee
|\Delta f\rangle\equiv\ket{f_a}-\ket{f_b}=-\mathcal{N}_a^{-1}\ket{\alpha}v_\text{coll}.
\label{deltaForce}
\eee
From (\ref{Velocity}) one can also obtain the discontinuous change in the velocity field induced by the collision
\begin{equation}
|\Delta V\rangle\equiv| V_a\rangle-| V_b\rangle=-\mathcal{S}_a^{t}\mathcal{N}_a^{-1}\ket{\alpha}v_\text{coll}.
\label{DeltaV}
\end{equation}
In Fig.~\ref{collfig1}b and \ref{collfig1}c we show examples of $|\Delta f\rangle$ and $|\Delta V\rangle$, respectively. By construction, the force in contact $\alpha$ before the collision is zero while the force in $\alpha$ after the collision is precisely $f_\text{coll}$, hence, 
\bee
f_\text{coll}= \langle\alpha|\Delta f\rangle=- \Omega_vv_\text{coll}
\label{collForce}
\eee
where 
\begin{equation}
\Omega_v\equiv  \bra{\alpha}\mathcal{N}_a^{-1}\ket{\alpha}.
\label{Omegav}
\end{equation}
In Fig.\ref{collfig2}c the predicted scaling law (\ref{collForce}) is shown to be in very good agreement with our numerics. Notice that $\Omega_v\equiv \frac{\bra{\Delta V}\Delta V\rangle}{v_\text{coll}^2}$, and can thus be interpreted as the volume where the change on the particles' velocities is of order of the velocity fluctuations $V_{na}$. In the following section we will show that a single correlation volume exist, hence $\Omega_v\sim\Omega_c\sim1/\delta z$. 

\section{Correlation Volume}

%The change in the velocities (\ref{DeltaV})  can be understood as the response of the system to a perturbation in the contact $\alpha$ of magnitude $v_\text{coll}$. Such local perturbations, generated by collisions, are known to display an exponential decay from the source \cite{During13,During14} ensuring a well defined correlation volume.
 
\begin{figure}[h]
\includegraphics[width=1.1\columnwidth,trim=0.3cm 0cm 0cm 0cm, clip=true ]{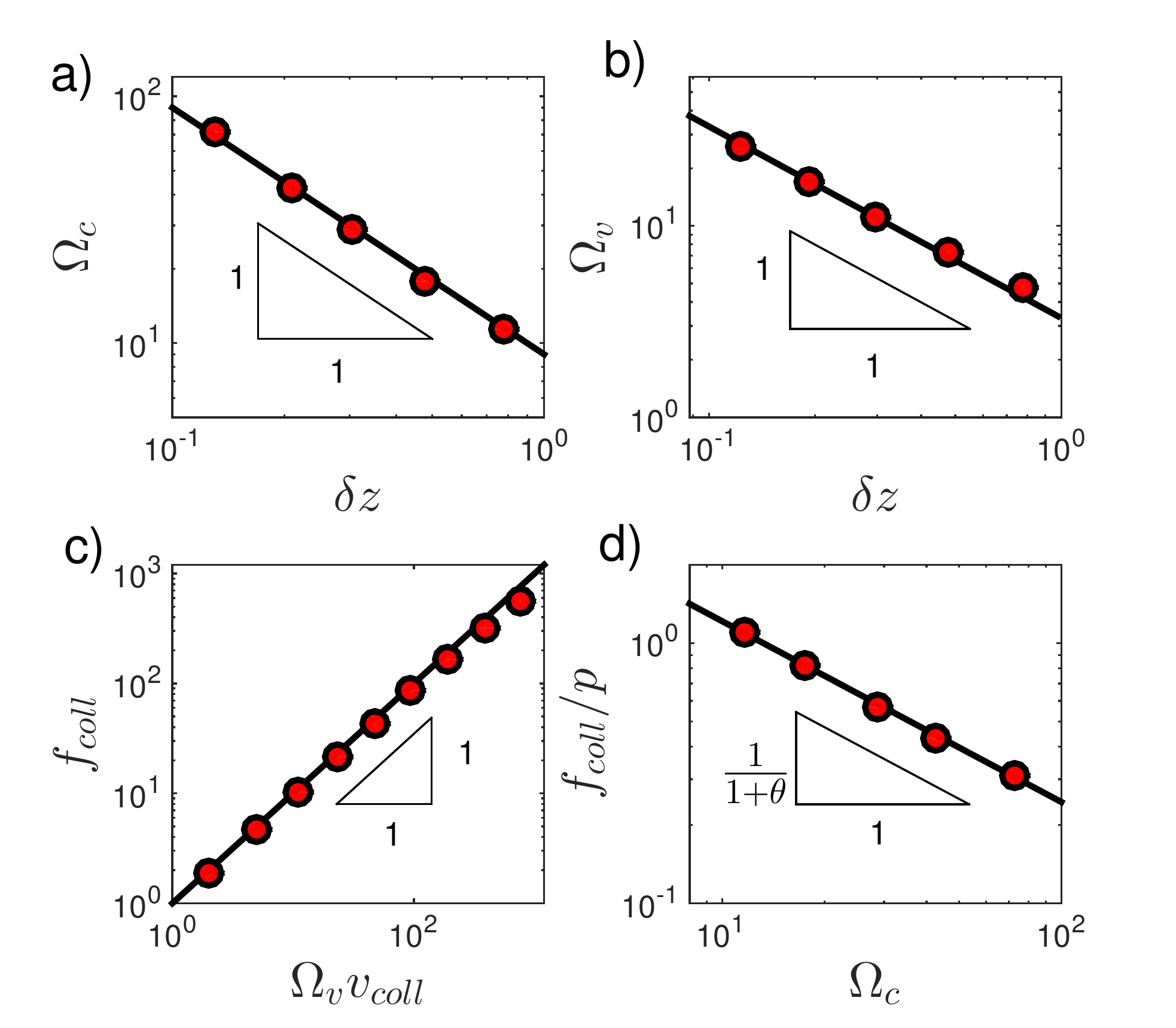} 
\caption{Data from simulations of the ASM in three dimensions, for systems of $N=2000$ particles, and pressures ranging between $10$ and $1000$. {\bf a)} The correlation volume $\Omega_c$  {\it vs} the coordination $\delta z$.   {\bf b)} The correlation volume $\Omega_v$  {\it vs} the coordination $\delta z$  {\bf c)} Mean contact force created at collisions $f_\text{coll}$ {\it vs} the product of the correlation volume $\Omega_v$ and the relative radial velocity at the collision $v_\text{coll}$. {\bf d)} Contact force created upon collisions normalized by the pressure $f_\text{coll}/p$ {\it vs} the correlation volume $\Omega_c$. The exponent $\theta=0.44$. }
\label{collfig2}
\end{figure}

The correlation volume $\Omega_v$ can be calculated using the spectral decomposition of the $\mathcal{N}_a$ matrix, with $\omega^2$ the eigenvalues and $\ket{r_\omega}$ the respective eigenmode.  From the definition of the correlation volume (\ref{Omegav}) one gets $\Omega_v=\sum_\omega \frac{|\langle \alpha | r_\omega \rangle|^2}{\omega^2}.$
The normalization of the eigenmodes implies that $\langle \alpha | r_\omega \rangle\sim~1/\sqrt{N_c}$, therefore in the thermodynamic limit
\begin{equation}
\Omega_v\sim \int \frac{D(\omega)}{\omega^2},
\label{OmegavD}
\end{equation}
where $D(\omega)$ is the eigenfrequencies distribution of the $\mathcal{N}$-matrix. The distribution $D(\omega)$ has been shown to display a plateau above a frequency scale $\omega^*\sim \delta z$, and up to frequencies $\omega\sim O(1)$ \cite{Lerner12a,During14} (modes below $\omega*$ are present but lead to sub-leading corrections in this argument). Thus one gets from Eq.(\ref{OmegavD}) that $\Omega_v\sim \frac{1}{\delta z}$, as shown in Fig.\ref{collfig2}b. Together with (\ref{collForce}) and (\ref{vcoll}), one gets from this result:
\bee
f_\text{coll}\sim \frac{\sqrt{p}}{\delta z}.
\label{fcollA}
\eee
The correlation volume:
\bee
 \Omega_c\equiv \frac{\langle \Delta f|\Delta f\rangle}{f_\text{coll}^2}\sim \delta z^2\bra{\alpha}\mathcal{N}_a^{-2}\ket{\alpha}
 \label{Omega2}
 \eee
can be calculated in a similar way:% using the spectral decomposition of the $\mathcal{N}_a$-matrix.
$$\Omega_c\sim \delta z^2 \sum_\omega \frac{|\langle \alpha | r_\omega \rangle|^2}{\omega^4}\sim\delta z^2\int_{\omega^*}^1  \frac{D(\omega)}{\omega^4}\sim 1/\delta z.$$ Thus both correlation volumes scale identicaly, i.e.: 
\bee
\Omega_c\sim\Omega_v\sim \frac{1}{\delta z}
\label{Omegaf}
\eee
as shown in Fig.\ref{collfig2}a. %The two dimensional case will be considered elsewhere.

\section{Weakest force in the volume $\Omega_c$}

It was recently shown that mechanical stability requires the distribution of contact forces in packings of frictionless spheres to vanish at small forces \cite{Wyart12}, as  observed in \cite{Lerner12}. There is one subtlety however: contacts at low forces can be decomposed in two types: some are mechanically isolated, whereas others are coupled mechanically to the rest of the system \cite{Lerner13a,Charbonneau15}. Only the later are relevant for the present argument. 
For those one finds: $$ P(f/p)\sim \left(\frac{f}{p}\right)^{\theta}$$ with $\theta\approx0.44$  \cite{Lerner13a}. This result can be derived in infinite dimension using the replica trick \cite{Charbonneau14}, yielding a similar result $\theta\approx0.42$ that appears to be correct in any dimensions.

Considering that the force distribution in flow must converge to that of jammed packings as the jamming point is approached, the minimum force $f_\text{min}$ can be easily estimated. Indeed the number of contact forces inside the correlation volume is proportional to $\Omega_c$. The weakest force $f_\text{min}$ can be estimated by the relation $\frac{1}{\Omega_c}\sim\int^{f_{\text{min}}/p}_0 P(b)db$, which leads to $\frac{f_\text{min}}{p}\sim  \Omega_c^{-\frac{1}{1+\theta}}$. Using the stationarity condition (\ref{sca1}) one gets
\begin{equation}
\frac{f_\text{coll}}{p}\sim  \Omega_c^{-\frac{1}{1+\theta}},
\label{fmin}
\end{equation}
which is in good agreement with the data shown in Fig.\ref{collfig2}d. %, and represents an indirect measurement for the weak force distribution in flow.

Finally, from the relations (\ref{fmin}),(\ref{fcollA}), (\ref{Omegaf}), one obtains scaling relations relating structure to rheological and dynamical properties:
\begin{eqnarray}
\label{rrr}
 \delta z&\sim&p^{-\frac{1+\theta}{(4+2\theta)}}\sim p^{-0.30}  \\\label{lll} 
 \Omega_c&\sim& p^{\frac{1+\theta}{4+2\theta}}\sim p^{0.30},\\\label{ggg}
f_\text{coll}&\sim& p^{\frac{3+2\theta}{4+2\theta}}\sim p^{0.80}.
\label{res1}
\end{eqnarray} 
 %{\color{red}[Shall we show data for these scalings?]}
Eq.(\ref{rrr}) was tested numerically in \cite{DeGiuli15a}. Eq.(\ref{rrr})  combined with Fig.\ref{collfig2}a. verifies Eq.(\ref{lll}) and the latter equation combined with Fig.\ref{collfig2}d verifies Eq.(\ref{ggg}).

\section{Strain scale between collision in $\Omega_c$}
\label{StrainScaleSection}
The strain $\Delta \gamma_c$ between two consecutive collision in a volume $\Omega_c$,  can be obtained from the stationarity of the shear stress in the steady flow state. The  increase in the shear stress $\Delta \sigma$ induced by a collision that forms  some contact $\alpha$ can be computed from Eq.(\ref{sigma}) and Eq.(\ref{deltaForce}). It reads
\bee
\Delta \sigma= -\frac{\langle\gamma|\Delta f\rangle}{\Omega}=v_\text{coll}\frac{\langle\gamma|\mathcal{N}_a^{-1}|\alpha\rangle}{\Omega}.
\eee
Before the formation of the contact $\alpha$, no forces are exerted between those particles, hence from Eq.~(\ref{deltaForce}) one gets $\langle\alpha| \mathcal{N}_a^{-1}\ket{\alpha}v_\text{coll}=\langle\alpha| \mathcal{N}_a^{-1}\ket{\gamma}$.
The jump in the shear stress then scales as $\Delta\sigma\sim\frac{v^2_\text{coll}\Omega_v}{\Omega}\sim\frac{p}{\delta z\, \Omega}$. The stress relaxes between collisions, following $\frac{d \sigma}{d\gamma}\sim -\sigma^2\sim -p^2$, as shown in \cite{During14}. Stationarity then implies that the strain scale between two consecutive collision in the entire system must scale as $ \Delta\sigma \big{/} \frac{d \sigma}{d\gamma}\sim  \frac{1}{\delta z \,p\, \Omega}$. Therefore, the strain scale $\Delta\gamma_c$ between two consecutive collision in a volume $\Omega_c$ is given by  
\bee
\Delta\gamma_c\sim\frac{1}{\delta z \,p\, \Omega_c}\sim  \frac{1}{p}.
\label{gamac}
\eee

A collision induces a change in the velocity field, inside the correlation volume $\Omega_c$, of the order of the non affine velocity. Thus, it is expected that the velocity correlations (with respect to strain) start to decorrelate precisely at a strain scale on the order of $\Delta \gamma_c$ (See suplementary information), as observed in \cite{Olsson10a,During14}.

\section{Recovering marginal stability in flow}

Stationarity also imposes a constraint between the particles' displacements and the gaps between particles in suspension flows. Close to the jamming point the relative velocity between particles scales as the non-affine velocities.  Then, the relative displacements that take place in a strain scale $\Delta \gamma_c$  between consecutive collisions inside the correlation volume $\Omega_c$ scales as $ \Delta \gamma_c V_{na}$. Such displacements must be of the same order as the minimal gap $h_{\text{min}}$ inside $\Omega_c$, i.e. 
\bee
h_{\text{min}}\equiv \min_{\in\Omega_c} h\sim \Delta \gamma_c V_{na}.
\label{gap}
\eee

The gap distribution at scales smaller than $\Omega_c$ is expected to be the same  as for  jammed packing. The distribution of jammed packing at small gaps is known to rise as a power law $P(h)\sim h^\nu$ with $\nu\approx 0.38$ \cite{Lerner12,Charbonneau12,Lerner13a}. Then, the minimal gap inside a volume $\Omega_c$ is given by the relation $\frac{1}{\Omega_c}\sim~\int^{h_{\text{min}}}_0 h^{-\nu}dh$, from which we obtain $h_{\text{min}}\sim \Omega_c^{-\frac{1}{1-\nu}}$. Using this relation together with Eqs.~(\ref{Vna},\ref{Omegaf},\ref{gap},\ref{gamac}),  one gets 
\bee
p\sim\delta z^{-\frac{2}{1-\nu}},
\label{micro2}
\eee 
which is a second, independently-derived expression that connects the suspension's macroscopic pressure with its microstructure. Comparing equations (\ref{micro2}) and (\ref{res1}), one finds that the exponents $\theta$ and $\nu$ must be related by  $\frac{1}{1-\nu}=\frac{2+\theta}{1+\theta}$. This relation between exponents was previously established for jammed packings, and was shown to be a consequence of their intrinsic marginal stability \cite{Wyart12,Lerner13a}. The extension of this relation below the jamming critical point can be interpret as follows: suspension flows remain  ``marginal stable'' far from the jamming point at  scales smaller than $\Omega_c$.

\section{Discussion and Conclusion}

We have formulated a microscopic scaling theory for dense non-Brownian suspension rheology in the framework of the Affine Solvent Model. We build upon the stationarity of the collisional processes in steady flow states to establish several scaling relations between the pressure, coordination, strain-scales, and correlation volumes. The constitutive relations, known as the friction and dilatancy laws, can be derived via finite size scaling arguments and the assumption of perturbation around a jammed solid \cite{DeGiuli15a}. Obtaining them in the present approach that focuses on collisions would be very interesting.

 In previous works \cite{During13,During14} we showed that local perturbations, as well as velocity correlations, decay exponentially at distance $r > \xi\sim \frac{1}{\sqrt{\delta z}}$. However, the effective volume affected by a local perturbation we computed here is given by $\Omega_c\sim 1/\delta z$ which is much smaller (except in two dimension) than  the naive correlation volume given by $\xi^d$. There is no contradiction:  it simply signals that the leading term of the response to a contact forming decays  with distance $r$ as $\delta R^2_\alpha(r)\sim f(r/l_c)/r^{d-2}$, where $f(x)$ is a rapidly decaying function of its argument.

Finally, a central question is how universal  the present results are. First, we expect our results on the spatial effects of collisions to hold true when {\it inertia} is present. Indeed in the unified description of viscous flows and inertial flows of frictionless particles we proposed in  \cite{DeGiuli15a},  the properties of the contact network (which control collisions) are essentially the same in these two cases. Second, and most importantly, we also expect that in the suspension case, both our results on collisions as well as  those of \cite{DeGiuli15a} hold true when particles are {\it frictional}.   This is not obvious at all, because in the inertial case friction affects the scaling exponents near jamming \cite{Peyneau08,Radjai05}. However in the presence of inertia, the change of scaling behavior stems from a change in the dominant dissipation mechanism, which becomes dominated by friction instead of collisions close to jamming \cite{DeGiuli15b}. However in suspensions friction never seem to dominate dissipation, at least in the range probed by numerics \cite{Trulsson12}  and experiments \cite{Dagois15}. We thus expect our results to hold in real materials, where they could be tested via imaging with sufficient temporal and spatial resolution.

We thanks Eric DeGiuli, L. Yan, J. Lin and M. Battalia  for discussions. G.D. acknowledges support from FONDECYT Grant No. 1150463. E.L. acknowledges funding from the Amsterdam Academic Alliance fellowship. 
\bibliography{Wyartbibnew}

%\bibliography{reference8}{}

\vskip2cm
\newpage
\thispagestyle{empty}
\newpage

\section{Appendix}

\subsection{The affine solvent model}
\label{model}
The ASM is an idealized suspension model which considers $N$ frictionless hard spherical particles in a volume $\Omega$, immersed in a viscous fluid, and hydrodynamic interactions are neglected. The dynamics is overdamped and the viscous drag force is proportional to the velocity difference between the particles and the fluid velocity. The drag force (\ref{dragForce}) can be written in compact form using the bracket notation: 
\bee
|F\rangle=-\eta_0r_0(|V\rangle-|V^f\rangle)
\label{dragDef}
\eee
In addition to the drag force, particles in contact interact via repulsive contact forces. The force acting over particle $k$ due to particle $i$ is given by $f_{ik}\vec{n}_{ik}$ where $f_{ik}$ represents the amplitude of the contact force (taken to be positive), and $\vec{n}_{ik}$ points along the difference vector $\vec{R}_k - \vec{R}_i$. Since we consider overdamped dynamics, the drag force on each particle is balanced at all times by the contact forces exerted by other particles, hence 
\bee
\vec{F}_k+\sum_{i\neq k} f_{ik}\vec{n}_{ik}=0.
\label{forceBalance}
\eee
Interaction forces in hard sphere systems are different from zero only for particles which are in contact. Thus the sum in (\ref{forceBalance}) runs only over the particles in contact, which can be written in compact notation using the transpose of the  $\mathcal{S}$-operator \cite{Lerner13}:
\bee
|F\rangle+\mathcal{S}^t|f\rangle=0.
\label{hc2A}
\eee
Operating with the $\mathcal{S}$ matrix on both side of the above equation, and using expression (\ref{dragDef}) one finds
$$-\eta_0r_0\mathcal{S}|V\rangle+\eta_0r_0\mathcal{S}|V^f\rangle+\mathcal{S}\mathcal{S}^t|f\rangle=0.$$
The constraints imposed by the hard spheres as described by Eq.~(\ref{hc1}) imply that the first term in the above equation vanishes. Defining the matrix $\mathcal{N}=\mathcal{S}\mathcal{S}^t$ and denoting the relative radial velocity induced by the fluid as $|v^f\rangle=\mathcal{S}|V^f\rangle$, the fundamental equation of the ASM for the contact forces is obtained as
\bee
|f\rangle=-\eta_0r_0\mathcal{N}^{-1}|v^f\rangle.
\label{contactForceA}
\eee 
For a simple shear flow $|v^f\rangle=\dot{\gamma}|\gamma\rangle$ where  the components of the vector $|\gamma\rangle$ are given by $||\vec{R}_k-\vec{R}_i|| (\vec{n}_{ik}\cdot \vec{e}_x) (\vec{n}_{ik}\cdot \vec{e}_y)$. Together with the contact forces $|f\rangle$, we determine the key rheological observables of the suspension, and in particular:
\ba
\label{dragA}
\text{\bf drag force}&& \quad |F\rangle=\eta_0r_0\mathcal{S}^t\mathcal{N}^{-1}|v^f\rangle, \\
\label{velocityA}
\text{\bf velocity}&& \quad |V\rangle=-\mathcal{S}^t\mathcal{N}^{-1}|v^f\rangle+|V^f\rangle,\\
\label{pressureA}
\text{\bf pressure}&& \quad p\equiv \frac{\langle r|f\rangle}{D\Omega}=-\eta_0r_0 \frac{\langle r| \mathcal{N}^{-1}|v^f\rangle}{D\Omega},\\
\label{shearA}
\text{\bf shear stress}&& \quad \sigma\equiv-\frac{\langle \gamma|f\rangle}{\Omega}=\eta_0r_0\frac{\langle \gamma| \mathcal{N}^{-1}|v^f\rangle}{\Omega}.
\ea
 
A simple shear velocity profile preserves the packing fraction $\phi$, while the pressure fluctuates around some mean value in steady-state flows. Such fluctuations can become very large close to the jamming point due to finite size effects. In some situation is therefore advantageous to consider a constant pressure system in which the packing fraction fluctuates around some mean. This can be done in the ASM framework by allowing the system to dilate and contract in addition to the simple shear velocity profile \cite{Lerner13}. The relative radial velocity of the fluid is then given by $|v^f\rangle=\dot{\gamma}(|\gamma\rangle+\kappa|r\rangle)$ where $\kappa$ is the dilatancy per unit shear, the latter is determined by imposing a constant pressure in eq.\ref{pressureA}. The result is $$\kappa=\frac{pD\Omega/\dot{\gamma}\eta_0r_0-\langle r|\mathcal{N}^{-1}|\gamma\rangle}{\langle r|\mathcal{N}^{-1}|r\rangle}.$$ 
The constitutive equations as well as the bulk properties should not depend on the ensemble considered, whether the constant pressure or constant packing fraction ensemble. Nevertheless, some properties might depend on the nature of the boundary conditions such as the fluctuations or relaxation of global quantities. In this work, unless otherwise stated, the results are valid in both cases.

\subsection{Force change induced by particles collision}
\label{collision}

As stated in the main text, operating with the post-collisional $\mathcal{S}_a$-matrix on the pre-collisional velocities $\ket{V_b}$, one obtains 
\bee
\mathcal{S}_a|V_b\rangle= v_\text{coll}|\alpha\rangle,
\label{beforeCollA}
\eee
where $\alpha$ is the contact created in the collision. Since the force between a pair of particles that are not in contact is zero, the force balance condition (\ref{hc2A}) can be rewritten as $$|F_b\rangle+\mathcal{S}_a^t|f_b\rangle=0,$$ where $\bra{\alpha}f_b\rangle=0$.  Operating on both sides of the above equation by $\mathcal{S}_a$ and using the drag force definition Eq.\ref{dragForce}, one finds  $$-\mathcal{S}_a|V_b\rangle+\mathcal{S}_a|V^f\rangle+\mathcal{S}_a\mathcal{S}_a^t|f_b\rangle=0,$$ which is similar to the expression found in the last section. Notice that $\eta_0$ and $r_0$ were set to unity. Replacing Eq.\ref{beforeCollA} in the above equation leads to the expression  used in the main text 
\begin{eqnarray}
\ket{f_b}&=& - \mathcal{N}_a^{-1}\ket{v^f}+\mathcal{N}_a^{-1}\ket{\alpha}v_\text{coll},
\label{forceBeforeA}
\end{eqnarray}
where $\mathcal{N}_a=\mathcal{S}_a\mathcal{S}_a^t$ and $\ket{v^f}=\mathcal{S}_a|V^f\rangle$.
 
\subsection{Decorrelation strain scale}

We define the non-affine velocity correlation function as
 $$C(\gamma_0,\gamma)=\frac{\bra{V^0_\text{n.a}}V_\text{n.a}\rangle}{\sqrt{\bra{V^0_\text{n.a}}V^0_\text{n.a}\rangle\bra{V_\text{n.a}}V_\text{n.a}\rangle}},$$  where $V^0_\text{n.a}$ and $V_\text{n.a}$ denote the non-affine velocities at the strains $\gamma_0$ and $\gamma$ respectively. We aim at determining the difference $\Delta C= C(\gamma_0,\gamma+\delta\gamma)-C(\gamma_0,\gamma)$ which can be written as 
 \begin{eqnarray}
 \Delta C=C(\gamma_0,\gamma)\left(\frac{(1+\frac{\bra{V^0_\text{n.a}}\Delta V\rangle}{\bra{V^0_\text{n.a}}V_\text{n.a}\rangle})}{\sqrt{(1+\frac{2\bra{V_\text{n.a}}\Delta V\rangle}{\bra{V_\text{n.a}}V_\text{n.a}\rangle}+\frac{\bra{\Delta V}\Delta V\rangle}{\bra{V_\text{n.a}}V_\text{n.a}\rangle})}}-1\right)\nonumber\,,
 \end{eqnarray}
 where the velocity field at $\gamma+\delta\gamma$ is given by $| V_\text{n.a}\rangle+|\Delta V\rangle$. In Sect.~\ref{StrainScaleSection} we showed that the strain scale between collisions is given by $\delta \gamma=\frac{1}{\delta z p \Omega}$. We thus estimate the change in the velocity field as the change induced by a collision (\ref{DeltaV}) plus the change of the velocities in between collisions. Between collisions the velocities vary smoothly, hence the change in the velocity field between collisions is approximately given by  $|\partial_\gamma V(\gamma)\rangle \delta \gamma$. Thus to the lowest order in $\delta\gamma$ the correlation function can be approximated as
  \begin{eqnarray}
 \Delta C\approx C(\gamma_0,\gamma)\left(\frac{\bra{V^0_\text{n.a}}\Delta V\rangle}{\bra{V^0_\text{n.a}}V_\text{n.a}\rangle}-\frac{v_\text{coll}^2}{2}\frac{\bra{\alpha}\mathcal{N}_a^{-1}|\alpha\rangle}{\bra{V_\text{n.a}}V_\text{n.a}\rangle}\right). \label{foo}
 \end{eqnarray}
In the last expression we used that $\bra{V_\text{n.a}}\Delta V\rangle=0$ in a collision, as can be shown using Eq.~(\ref{DeltaV}), and we assume that $\bra{V_\text{n.a}}\partial_\gamma V\rangle\sim N^\nu$ with $\nu<1$. In general the scaling properties of $\frac{\bra{V^0_\text{n.a}}\Delta V\rangle}{\bra{V^0_\text{n.a}}V_\text{n.a}\rangle}$ are unknown. However, while the correlation function $C(\gamma_0,\gamma)\sim 1$ the velocity $V^0_\text{n.a}$ can be approximated by $V_\text{n.a}$, and the first term on the LHS of Eq.~(\ref{foo}) can be neglected. We finally find that the initial evolution of the correlation function is given by 
 \begin{eqnarray}
 \Delta C\approx -\frac{v_\text{coll}^2}{2}\frac{\bra{\alpha}\mathcal{N}_a^{-1}|\alpha\rangle}{\bra{V_\text{n.a}}V_\text{n.a}\rangle}C(\gamma_0,\gamma)\sim -\frac{1}{\delta z N}C(\gamma_0,\gamma). \nonumber
 \end{eqnarray}
 Since $\delta \gamma=\frac{1}{\delta z p \Omega} \sim \frac{1}{\delta z p N} $ one can rewrite the last expression as 
 $$ \Delta C\sim- p C(\gamma_0,\gamma) \delta \gamma.$$ 
In the limit of large $N$ the above expression represents a differential equation, the solution to which displays an exponential decay with strain over a decorrelation strain scale of $\frac{1}{p}$.

\end{document}